\documentclass[11pt,preprint,amsmath,amssymb,aps,showkeys,showpacs,nofootinbib]{revtex4}

\usepackage{soul}

\usepackage[colorlinks=true, pdfstartview=FitV, linkcolor=blue, citecolor=red, urlcolor=magenta]{hyperref}

\usepackage{graphicx}
\usepackage{latexsym}
\usepackage{amsmath}
\usepackage{amsfonts}
\usepackage{amssymb}
\usepackage{verbatim}
\usepackage{dcolumn}
\usepackage{amssymb}
\usepackage{bm}
\usepackage{float}
\usepackage{subfigure}
\usepackage[english]{babel}

\usepackage{todonotes}
\usepackage{booktabs}
\usepackage{multirow}

\begin{document}
\newcommand{\be}{\begin{equation}}
\newcommand{\ee}{\end{equation}}
\newcommand{\bea}{\begin{eqnarray}}
\newcommand{\eea}{\end{eqnarray}}
\newcommand{\hf}{\frac{1}{2}}
\newcommand{\pa}{\partial}

\newcommand{\NITK}{
\affiliation{Department of Physics, National Institute of Technology Karnataka, Surathkal  575 025, India}
}

\newcommand{\Shreyasusa}{\affiliation {Department of Oral Health Sciences, School of Dentistry,\\
University of Washington, Seattle, WA 98195, USA.
}}
\newcommand{\naveenbgu}{\affiliation {Department of Physics, Ben-Gurion University of the Negev,\\ Beer-Sheva 84105, Israel}}

\newcommand{\naveencroatia}{\affiliation {Theoretical Physics Division, Rudjer Bo\v skovi\'c Institute,\\  Bijeni\v cka c.54,  HR-10002 Zagreb, Croatia}}

\title{Spinning LQG black hole as a particle accelerator}

\author{Ullas P. Suresh}
\email{ullaspullursuresh@gmail.com}
\NITK

\author{Karthik R}
\email{raokarthik1198@gmail.com}
\NITK

\author{K. M. Ajith}
\email{ajith@nitk.ac.in}
\NITK

\author{Kartheek Hegde}
\email{hegde.kartheek@gmail.com}
\NITK

\author{Shreyas Punacha}
\email{shreyasp444@gmail.com, shreyas4@uw.edu}
\Shreyasusa

\author{A. Naveena Kumara}
\email{naviphysics@gmail.com, nathith@irb.hr}

\naveencroatia

\begin{abstract}
We demonstrate that the spinning LQG black hole can act as a cosmic particle accelerator. The LQG solution is singularity-free and can possess spin greater than that of a Kerr black hole. The additional black hole hair, arising from quantum effects, significantly influences the particle dynamics around the black hole. Under suitable physical conditions, the center-of-mass energy can grow arbitrarily high during the collision of two generic particles in the spacetime of an extremal black hole. In the non-extremal case, there exists a finite upper bound on the center-of-mass energy, the maximum value of which depends on the LQG parameter. These results are particularly interesting from an astrophysical perspective, especially in the context of probing Planck-scale physics.
\end{abstract}


\maketitle

\section{Introduction} \label{sec_intro}

Since their inception, black holes have been among the most intriguing objects in physics. While the theoretical aspects of black holes have been extensively explored for a long time, the observational confirmation and validation of these theoretical predictions have only been achieved recently. The LIGO-Virgo-KAGRA collaboration's detection of gravitational waves originating from the collision of compact binary objects provided the first direct evidence for the existence of black holes, opening a new field of research known as black hole spectroscopy \cite{Abbott:2016blz, TheLIGOScientific:2016src, Abbott:2016nmj}.  Furthermore, the Event Horizon Telescope (EHT) succeeded in imaging the shadow of a black hole, an achievement made possible through the observation of gravitational light bending and photon capture at the event horizons of supermassive black holes in the center of the giant elliptical galaxy M87$^\star$ and at the center of our galaxy, Sgr A$^\star$ \cite{Akiyama:2019bqs, Akiyama:2019cqa, Akiyama:2019fyp, EventHorizonTelescope:2022wkp, EventHorizonTelescope:2022xqj} \footnote{A novel method is the Gaussian bending which could serve as a tool for testing various gravity theories and distinguishing physical objects with singularities from those without \cite{Zhang:2024uex, Zhang:2021ygh}}. Although these observations align with the predictions of general relativity with remarkable precision, fundamental questions remain unresolved. One such question pertains to the validity of the cosmic censorship conjecture, which says that spacetime singularities are always hidden behind an event horizon, thereby preventing the existence of naked singularities. In the context of black hole physics, these questions are of significant importance, as black holes are strongly gravitating objects that may possess spacetime singularities at their centers. However, the combination of gravity with quantum theory could potentially mitigate the occurrence of curvature singularities in the Universe.

It is well known that Loop Quantum Gravity (LQG) is a candidate for quantum gravity, effectively resolving singularity problems not only in black hole spacetimes \cite{Ashtekar:2005qt, Modesto:2005zm, Boehmer:2007ket, Campiglia:2007pb, Gambini:2008dy} but also in cosmological models, particularly by eliminating the Big Bang singularity in Loop Quantum Cosmology \cite{Ashtekar:2006wn, Ashtekar:2006es, Vandersloot:2006ws}. One of the most effective methods for addressing black hole singularities in LQG is phase space quantization, or semi-classical polymerization, which preserves the inherent discreteness of spacetime predicted by LQG \cite{Boehmer:2007ket, Campiglia:2007pb, Gambini:2008dy}. In this context, various polymerization schemes have been proposed, leading to different types of regular black hole spacetimes. In Ref. \cite{Peltola:2009jm}, it was shown that using effective polymerization schemes, the interior of a D-dimensional spherically symmetric black hole can evolve into a complete, regular, single-horizon spacetime where the classical singularity is replaced by a bounce. Furthermore, the semi-classical polymerization of the interior of Schwarzschild black holes results in a non-singular, single-horizon black hole spacetime that is asymptotically flat \cite{Peltola:2009jm, Peltola:2008pa}. 

Spinning black hole solutions are of significant interest because most astrophysical black holes are characterized by definite angular momentum \cite{EventHorizonTelescope:2022xqj}. However, it should be noted that studies of black hole singularities in LQG are mostly confined to spherically symmetric spacetimes due to their relative simplicity. Nevertheless, a spinning LQG black hole solution was constructed in Ref. \cite{Kumar:2022vfg} using a modified Newman-Janis algorithm. In addition to the conventional black hole parameters, such as mass and angular momentum, the spinning LQG Kerr-like solution introduces an additional parameter, $\ell$, arising from quantum effects. It was shown that this additional parameter can be constrained by studying the shadow cast by M87$^\star$ and Sgr A$^\star$ as observed by the Event Horizon Telescope \cite{Afrin:2022ztr, Islam:2022wck, Kumar:2023jgh, Vagnozzi:2022moj}.

A notable feature of spinning black holes is the possibility of extracting mechanical energy from them. The concept of extracting the rotational energy of a black hole, originally introduced by Penrose, is known as the \emph{Penrose process} \cite{Penrose:1971uk}. In this process, under certain physical conditions, a particle scattered by a black hole can emerge with more energy than it originally had, with the extra energy derived from the spin of the black hole. This process is particularly efficient in scenarios where two particles collide near the event horizon of the black hole, producing two or more resultant particles. One of these particles may escape, carrying more energy than the incoming ones; this phenomenon is referred to as the \emph{collisional Penrose process} \cite{Piran1977}. Interest in this concept was rejuvenated with the introduction of the Bañados-Silk-West (BSW) mechanism \cite{Banados:2009pr}. This mechanism suggests that rotating black holes can function as natural particle accelerators, potentially explaining highly energetic astrophysical phenomena such as active galactic nuclei, gamma-ray bursts, and ultra-high-energy cosmic rays. The BSW mechanism significantly increases the efficiency of energy extraction, potentially reaching energy levels on the order of Planck-scale physics \cite{Bejger:2012yb, Schnittman:2014zsa, Berti:2014lva, Leiderschneider:2015kwa}. In the BSW mechanism, the center-of-mass energy of colliding particles can become exceedingly high near the event horizon of an extremal black hole, particularly when one particle has critical angular momentum. Due to its theoretical and observational importance, this mechanism has been the subject of extensive research \cite{Berti:2009bk, Banados:2010kn, Jacobson:2009zg, Zaslavskii:2010aw, Wei:2010gq, Lake:2010bq, Wei:2010vca, Liu:2010ja, Mao:2010di, Zhu:2011ae, Zaslavskii:2012fh, Zaslavskii:2012qy, Zaslavskii:2010pw, Grib:2010xj, Harada:2011xz, Liu:2011wv, Patil:2010nt, Patil:2011aw, Patil:2011ya, Patil:2011uf, Amir:2015pja, Ghosh:2014mea, NaveenaKumara:2020kpz, AhmedRizwan:2020sza,  Nosirov:2024vsu, Galajinsky:2013as, Tsukamoto:2013dna, Ogasawara:2018gni} (see Ref. \cite{Harada:2014vka} for a brief review).

This article focuses on investigating particle collisions near a spinning LQG black hole. The main motivation for this study stems from the observation that this new solution enables the black hole to possess a higher spin than Kerr and Kerr-Newman black holes, which may improve the effectiveness of extracting rotational energy. Additionally, unlike the Kerr black hole, this solution is singularity-free. It also introduces modifications to the horizon structure, where the LQG parameter affects the size of the event horizon and the static limit surface. As we will demonstrate, the LQG parameter has a significant impact on the BSW mechanism. The article is organized as follows. In section \ref{sec1}, we present the details of the black hole solution. In section \ref{sec2}, we study the particle motion in that black hole spacetime, followed by an analysis of particle collisions in section \ref{sec3}. We conclude the article in section \ref{sec4} with results and discussions.

\section{Spinning LQG black hole}\label{sec1}

In this section, we briefly present the details of the black hole spacetime. We consider the solution obtained by the authors in Ref. \cite{Kumar:2022vfg}, which describes LQG-inspired multihorizon rotating black holes derived from a partially polymerized static spherically symmetric black hole solution \cite{Peltola:2009jm}, using the Newman–Janis algorithm \cite{Azreg-Ainou:2014pra, Azreg-Ainou:2014aqa}. The explicit form of the metric is given by
\begin{equation} \label{metriceq}
\begin{aligned}
ds^2 = &- \left(1 - \frac{2M(r)\sqrt{r^2 + \ell^2}}{\rho^2}\right) dt^2 +  \frac{\rho^2}{\Delta} dr^2 + \rho^2 d\theta^2 - \frac{4aM(r)\sqrt{r^2 + \ell^2}\sin^2\theta}{\rho^2} dt d\varphi \\ &+ \frac{\mathcal{A}\sin^2\theta}{\rho^2} d\varphi^2 
\end{aligned}
\end{equation}
where
\begin{align}
    M(r) &= M - \frac{r - \sqrt{r^2 + \ell^2}}{2}, &  \rho^2 &= r^2 + \ell^2 + a^2\cos^2\theta, \\
    \Delta &= r^2 + \ell^2 + a^2 - 2M(r)\sqrt{r^2 + \ell^2}, & \mathcal{A} &= (r^2 + \ell^2 + a^2)^2 - a^2 \Delta \sin^2\theta.
\end{align}
Here, \( M \) is the black hole mass, which we set to \( M = 1 \) for simplicity wherever necessary. The spin parameter \( a \) is related to the angular momentum \( J \) and mass \( M \) of the black hole as \( a = J / M \), and \( \ell \) is the polymer length scale, a positive constant \footnote{The polymer length scale represents a fundamental minimal length in polymer quantization, analogous to the discrete structures in loop quantum gravity \cite{Ashtekar:2002sn}. This scale arises naturally from quantum geometry, highlighting the inherent discreteness of space, and plays a critical role in modifying the Hamiltonian of a system into a polymeric Hamiltonian.}.  In natural units, both $a$ and $\ell$ have the dimensions of length. The metric is asymptotically flat as $r \to \infty$. Additionally, the black hole solution exists for all values of $\ell$. Although the metric resembles the well-known Kerr black hole solution, it captures important features of LQG, such as a transition surface at the black hole center and the global regularity of spacetime. We present a brief overview of the spacetime properties, focusing mainly on the horizon structure and ergoregion. For further details on the spacetime structure, including the Penrose diagram, see Ref. \cite{Kumar:2022vfg}. Despite its complex spacetime structure, the metric possesses notable limiting cases. In the limit $\ell \to 0$, we recover the Kerr black hole spacetime; in the limit $a \to 0$, we obtain spherical LQG black holes; and when both $a$ and $\ell$ vanish, we recover the Schwarzschild spacetime. Thus, the polymer length scale $\ell$ is key to the rich spacetime structure.

\begin{figure}[ht!]
    \centering
    \includegraphics[width=\textwidth]{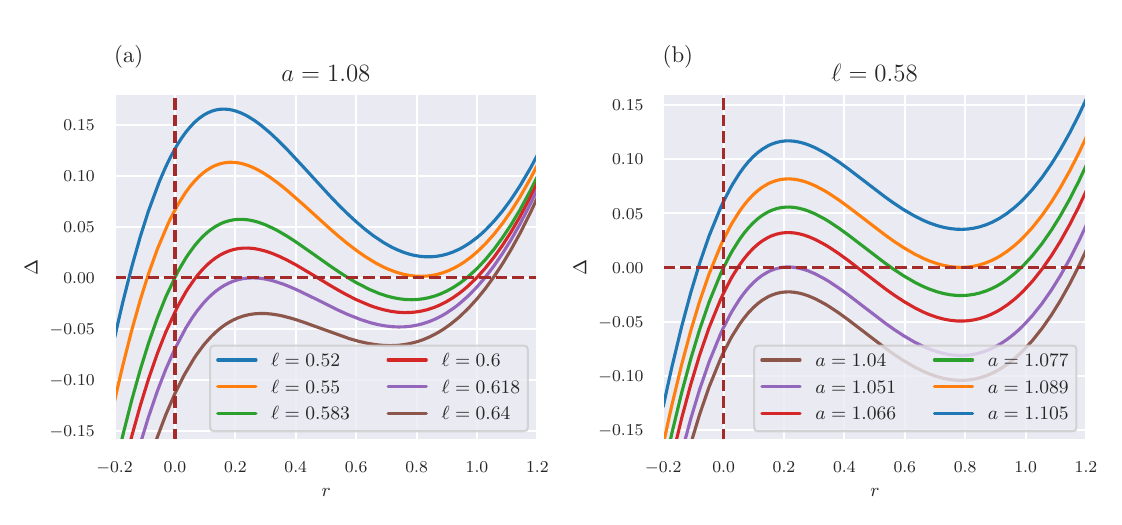}
    \caption{%
    The plots depicting the different horizon structures of spinning LQG black holes, given by the roots of $\Delta=0$. 
    Left: For a fixed $a$, varying $\ell$ changes $\Delta$ so that one, two, or three positive roots (i.e., horizons) may form, depending on whether $\Delta$ develops both a local maximum and minimum ($\ell < M/\sqrt{2}$) or remains strictly increasing ($\ell \ge M/\sqrt{2}$). 
    Right: For a fixed $\ell$, increasing $a$ similarly alters the horizon structure.}
    \label{fig_horizon}
\end{figure}

The metric has divergences at $\Delta = 0$, indicating a coordinate singularity corresponding to the black hole horizon. The location of this null surface is determined by the LQG parameter $\ell$ as well as the conventional parameters $a$ and $M$. The classical ring singularity of the Kerr black hole is replaced by a timelike transition surface at $r = 0$ inside the event horizon. Another contrasting feature to the Kerr spacetime is the multihorizon structure, as illustrated in Fig.~\ref{fig_horizon}. The function \(\Delta\) governs the horizon structure and can exhibit up to three positive roots when \(\ell < M/\sqrt{2}\), owing to a local maximum (at $r=(M-\sqrt{M^2-2\ell^2})/2$) and minimum (at $r=(M+\sqrt{M^2-2\ell^2})/2$). In this case, black holes may have multiple horizons. Conversely, for \(\ell \geq M/\sqrt{2}\), \(\Delta(r)\) is strictly increasing, admitting at most one horizon. The parameter space of $\ell$ and $a$ produces three distinct scenarios (excluding no-horizon spacetime): black holes with one ($r_1$), two ($r_1 > r_2$), or three horizons ($r_1 > r_2 > r_3$), where these are the positive roots of the condition $\Delta = 0$. Here, $r_1\equiv r_{\text{h}}$ is the event horizon, $r_2 \equiv r_{\text{c}}$, if it exists, is the Cauchy horizon, and $r_3$ (conditionally present) is an additional horizon inside the Cauchy horizon. For fixed values of $a$ or $\ell$, there exists a critical value of $\ell$ or $a$ for which the Cauchy horizon and event horizon merge, resulting in a maximally spinning black hole. This merging is characterized by the simultaneous solution of the equations $\Delta = 0$ and $\Delta' = 0$
\begin{equation}
\begin{split}
    \Delta &= a^2 + \sqrt{\ell^2 + r^2} (r - 2M) = 0, \\
    \Delta' &= \frac{\ell^2 + 2r(r - M)}{\sqrt{\ell^2 + r^2}} = 0.
\end{split}
\end{equation}
For a given $\ell$, we can calculate the extremal black hole's radius of the event horizon $r_\text{he}$ and the extremal spin $a_\text{e}$ as
\begin{equation}
\begin{split}
     r_{\text{he}} &= \frac{1}{2} \left(\sqrt{M^2 - 2\ell^2} + M\right), \\
     a_{\text{e}} &= \frac{1}{2^{3/4}} \left(\left( 3M - \sqrt{M^2 - 2\ell^2}\right) \sqrt{M \left(\sqrt{M^2 - 2\ell^2} + M\right) + \ell^2}\right)^{1/2}.
\end{split}
\end{equation}
In the limit $\ell \to 0$, both $a_{\text{e}}$ and $r_{\text{he}}$ become $M$, corresponding to the spin and event horizon radius of an extremal Kerr black hole, respectively. Interestingly, in some cases (not necessarily extremal), there exists a black hole solution with $a/M > 1$, which is constrained in the Kerr black hole case. It was shown that the maximum rotation of Kerr black holes cannot reach $M$ \cite{Thorne:1974ve}, but this constraint is relaxed here.

\begin{figure}[t!]
    \centering
    \includegraphics[width=\textwidth]{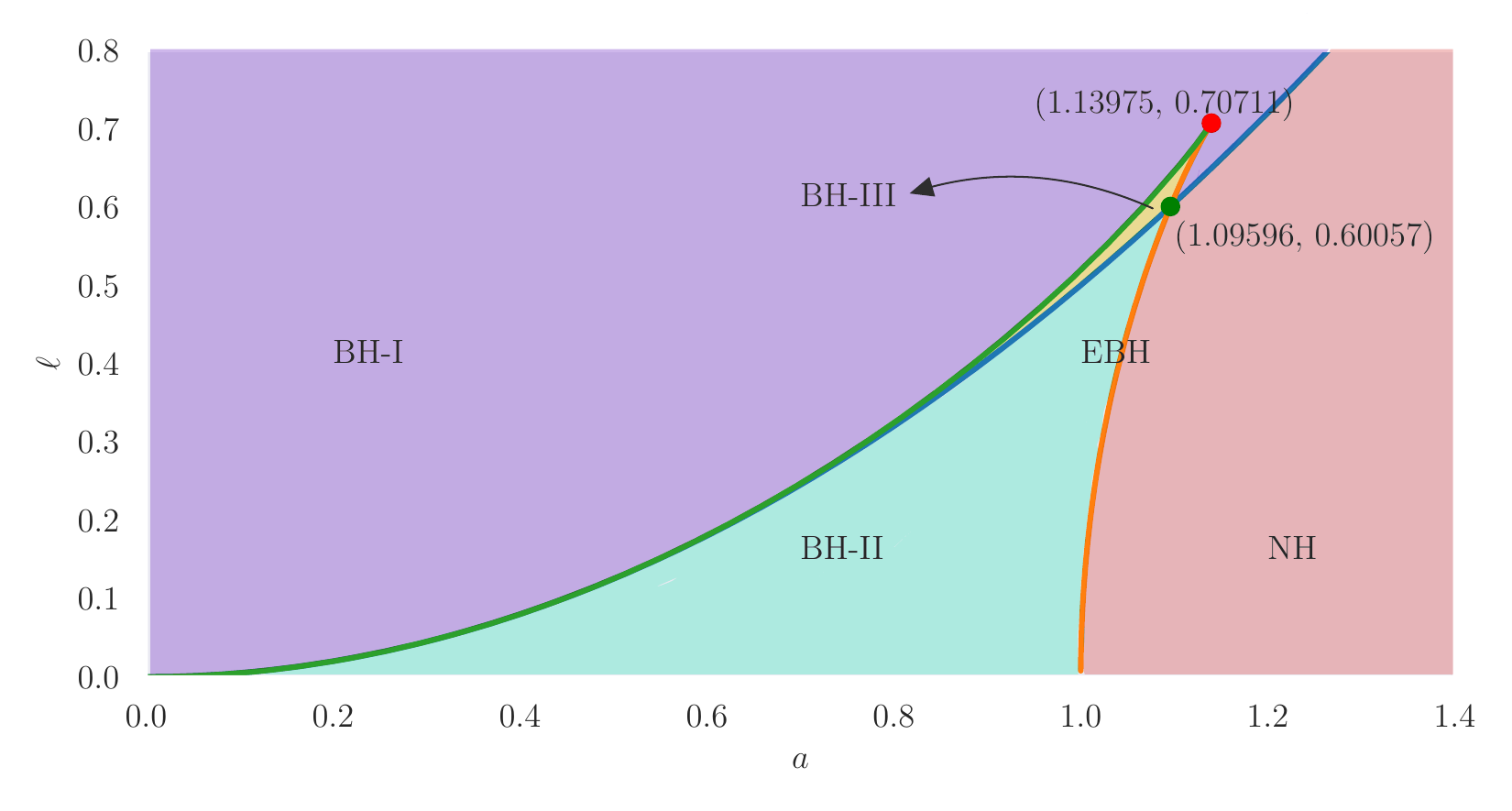}
    \caption{%
    The parameter space of $(a, \ell)$ for the spinning LQG black hole, illustrating regions of distinct horizon structures. 
    BH I: single-horizon black holes, 
    BH II: black holes with two horizons, 
    BH III: black holes with three horizons, 
    EBH: extremal black holes, 
    NH: no-horizon spacetimes. 
    The red curve indicates the critical $(a, \ell)$ values for which the event and Cauchy horizons coincide (extremal case), and beyond which there are no black hole solutions.
    The green line separates the BH I and BH III regions, while the blue line separates BH III from BH II and BH I from the NH regions.}
    \label{fig_parameter}
\end{figure}

The parameter-space diagram in Fig.~\ref{fig_parameter} further demonstrates how varying \((a, \ell)\) leads to distinct black hole (BH) solutions, ranging from single- or multi-horizon configurations (BH I, BH II, BH III) to no-horizon (NH) spacetimes, and includes the orange line of extremal black holes (EBH) where horizons merge. Although black hole solutions may exist for arbitrary values of the parameter $\ell$, extremal solutions are guaranteed only within a specific range of $\ell$ values (orange line). In the range $0 \leq \ell \leq 0.60057$, the scenario is analogous to Kerr spacetime, where one finds non-extremal black holes, extremal black holes, and no-horizon spacetimes depending on the value of $a$. Specifically, for a given value of $\ell$, a non-extremal black hole becomes extremal when $a = a_{\text{e}}$, and for $a > a_{\text{e}}$, the solutions correspond to no-horizon spacetimes. The extremal solution at $\ell = 0.60057$ represents an extremal black hole with a null throat (green dot in Fig. \ref{fig_parameter}), characterized by $(a_{\text{e}} = 1.09596, \ell = 0.60057)$. In the domain $0.60057 < \ell \leq 0.70711$, black hole solutions persist even for values of $a$ slightly greater than $a_{\text{e}}$, within a narrow region, before they transition into no-horizon spacetimes. The extremal black hole solution at $\ell _{\text{e}} = 0.70711$, called ultraextremal black hole (red dot in Fig. \ref{fig_parameter}), is characterized by $(a_{\text{e}} = 1.13975, \ell _{\text{e}} = 0.70711)$. For $\ell > 0.70711$, extremal black hole solutions cease to exist (see Ref. \cite{Kumar:2022vfg} for more details on this). Since the center-of-mass energy of colliding particles in the BSW mechanism increases indefinitely only in extremal cases, we restrict our consideration of the LQG parameter $\ell$ to the domain $0 \leq \ell \leq 0.70711$.

\begin{figure}[t!]
    \centering
    \includegraphics[width=\textwidth]{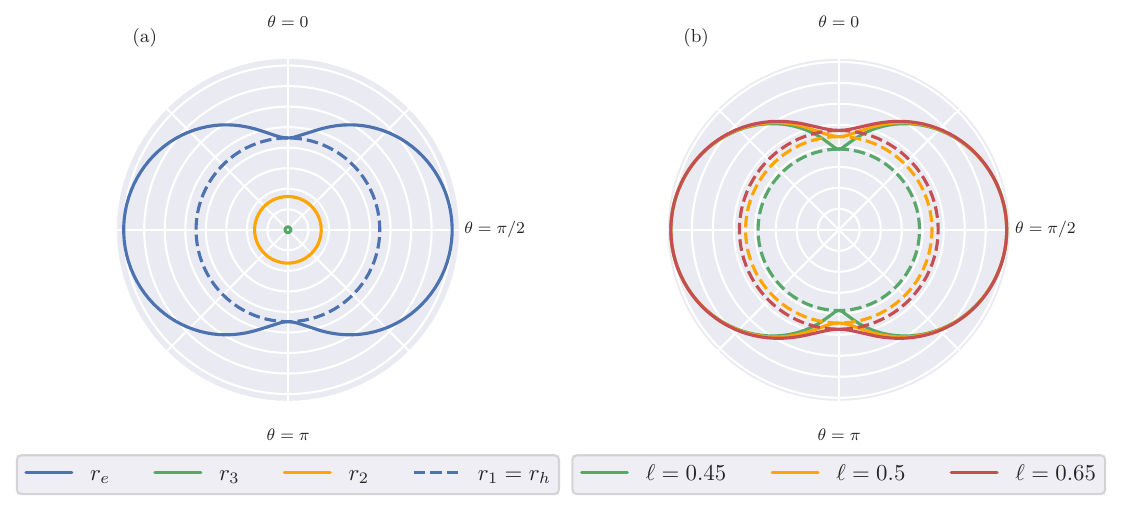}
    \caption{(a) The typical horizon structure of a spinning LQG black hole. Three generic horizons are possible, marked as $r_1$, $r_2$, and $r_3$, corresponding to the positive roots of $\Delta = 0$. The largest root, $r_1$, represents the event horizon. The cases with a single horizon and two horizons are discussed in the text. The outermost surface in this figure denotes the static limit surface, $r_\text{e}$. (b) The variations of the event horizon and static limit surface with the LQG parameter $\ell$. }
    \label{fig_ergo}
\end{figure}

The metric is stationary and axisymmetric with corresponding Killing vectors $t^a$ and $\varphi^a$. Recall that a Killing horizon is a null surface whose null generators coincide with the orbits of a one-parameter group of isometries. A Killing vector field is normal to the Killing horizon. The rigidity theorem states that the event horizon of a stationary black hole must be a Killing horizon \cite{Wald:1999vt}. Carter \cite{Carter:2009nex} showed that for a static black hole, the event horizon itself is a Killing horizon, whereas for a stationary axisymmetric spacetime, there exists a Killing field $\xi^a = t^a + \Omega \varphi^a$ normal to the event horizon, where $\Omega$ is the angular velocity of the horizon. Since Carter's theorem does not rely on specific field equations, its applicability to the metric \eqref{metriceq} is valid. The surface where $g_{tt}$ changes sign is a Killing surface associated with the Killing vector $t^a$, called the static limit surface. It lies outside the event horizon and extends depending on the angular coordinate $\theta$, in addition to the black hole parameters. It coincides with the event horizon at the poles. The region between this surface and the event horizon is called the ergoregion, where both the $t$ and $r$ coordinates are timelike. The spacetime in this region is dragged along the rotational direction of the black hole, known as the frame dragging effect, forcing any particle to move in this direction. However, particles can escape from this region, as the static limit surface is not a no-return surface like the event horizon.

An interesting phenomenon is that an ingoing particle can enter the static limit surface and escape to infinity with more energy than it initially had, harnessing the black hole's rotational energy. This possibility makes a spinning black hole a potential source of high-energy astrophysical phenomena such as gamma-ray bursts and active galactic nuclei. The volume of the ergoregion is influenced by the LQG parameter $\ell$, as both the ergoregion radius $r_{\text{e}}$ and the event horizon $r_{\text{h}}$ depend on it. We note that the extension of the ergosurface is $2M$ at the equatorial plane for all values of $\ell$, with variations due to $\ell$ appearing for other $\theta$ values, particularly significant near the poles, as depicted in Fig. \ref{fig_ergo}. The size of the ergoregion affects the rotational energy extraction process, deviating from that of a Kerr black hole. Details of the horizon structure and the ergoregion are presented in Table \ref{tab1}. In summary, the inner horizon structure is not crucial for investigating particle acceleration in the vicinity of the event horizon of the black hole. The primary concern of this study is the size of the ergoregion, which is determined by the event horizon and the static limit surface.

\begin{table}[h!]
\begin{tabular}{m{1em} m{5.6em} m{3.5em} m{3.5em} m{3.5em} m{1em} m{5em} m{5em}}
\multicolumn{1}{c}{$\ell$} & $a $  & $r_1\equiv r_{\text{h}}$ & $r_2 \equiv r_{\text{c}}$ & $r_3$ & $r_{\text{e}}$ & $\delta_1 = r_{\text{h}}-r_{\text{c}}$ & $\delta_2 = r_{\text{h}}-r_{\text{e}}$\\
\hline
\multicolumn{1}{c}{\multirow{3}{*}{0}} 
& $0.6 $ & 1.8  &  0.2  &  -  & 2  &  1.6  &  0.2   \\
& $0.8$  & 1.6  &  0.4  &  -  & 2  &  1.2  &  0.4   \\
& $a_{\text{e}} = 1$ & 1  &  1  &  -  & 2  &  0  &  1  \\
\hline
\multicolumn{1}{c}{\multirow{6}{*}{0.2}} 
& $0.6 $ & 1.80137  &  -  &  -  & 2  &  -  &  0.19863   \\
& $0.63085 $ & 1.77750  &  0.02041  &  0.02041  & 2  &  1.75709  &  0.22250   \\
& $0.631 $ & 1.77739  &  0.02669  & 0.01417  & 2  &  1.75069  &  0.22261   \\
& $0.8$  & 1.60408  &  0.32577  &  -  & 2  &  1.27831  &  0.39592   \\
& $1.0$ & 1.12432  &  0.83394  &  -  & 2  &  0.29038  &  0.87568\\
&$a_{\text{e}}= 1.01005$ & 0.97958  &  0.97958  &  -  & 2  &  0  &  1.02042   \\
\hline
\multicolumn{1}{c}{\multirow{6}{*}{0.4}} 
& $0.6 $ & 1.80530  &  -  &  -  & 2  &  -  &  0.19470   \\
& $0.8$  & 1.61543  &  -  &  -  & 2  &  -  &  0.38457   \\
& $0.88492$  & 1.49352  &  0.08768  &  0.08768  & 2  &  1.40584  &  0.50648   \\
& $0.885$  & 1.49339  &  0.09593 &  0.07951  & 2  &  1.13974  &  0.50661   \\
& $1.0$ & 1.22264  &  0.57898  &  -  & 2  &  0.64366  &  0.77736   \\
& $a_{\text{e}}= 1.04091$  & 0.91231  & 0.91231  &  -  & 2  &  0  &  1.08769   \\
\hline
\multicolumn{1}{c}{\multirow{6}{*}{0.6}} 
& $0.6 $ & 1.81133  &  -  &  -  & 2  &  -  &  0.18867   \\
& $0.8$  & 1.63191  &  -  &  -  & 2  &  -  &  0.36809   \\
& $1.0$  & 1.30281  &  -  &  -  & 2  &  -  &  0.69719   \\
& $1.06645$  & 1.07839  & 0.23542  &  0.23542  & 2  &  0.84297  &  0.92161   \\
& $1.075$  & 1.03177  &  0.41096 & 0.10033  & 2  &  0.62081  &  0.96822   \\
& $a_{\text{e}} = 1.09576$ & 0.76457  & 0.76457  &  -  & 2  &  0  &  1.23543   \\
\hline
\multicolumn{1}{c}{\multirow{4}{*}{0.60057}} 
& $0.6 $ & 1.81136  &  -  &  -  & 2  &  -  &  0.18864   \\
& $0.8$  & 1.63200  &  -  &  -  & 2  &  -  &  0.36800   \\
& $1.0$  & 1.30317  &  -  &  -  & 2  &  -  &  0.69683   \\
& $a_{\text{e}} = 1.09596$ & 0.76393  &  0.76393  &  0  & 2  &  0  &  1.23607   \\
\hline
\multicolumn{1}{c}{\multirow{4}{*}{0.70711}} 
& $0.6 $ & 1.81519  &  -  &  -  & 2  &  -  &  0.18481   \\
& $0.8$  & 1.64200  &  -  &  -  & 2  &  -  &  0.35800   \\
& $1.0$  & 1.33994  &  -  &  -  & 2  &  -  &  0.66006   \\
& $a_{\text{e}} = 1.13975$ & 0.5  &  0.5  &  0.5  & 2  &  0  &  1.5   \\
\end{tabular}
\caption{The horizon structure and static limit surface for the LQG rotating black hole for different values of $\ell$ and $a$. The static limit surface is shown in the equatorial plane, where it has maximum extension. } \label{tab1}
\end{table}

\section{Particle Motion around a Spinning LQG Black Hole}\label{sec2}

In this section, we study the particle motion around the rotating LQG black hole using Hamilton-Jacobi theory. The Lagrangian of a particle in the geodesic path is given by
\begin{equation}
 \mathcal{L} = g_{\mu\nu}\dot{x}^\mu \dot{x}^\nu 
\end{equation}
where the dot denotes differentiation with respect to an affine parameter \(\lambda\). The parameter \(\lambda\) is chosen such that \(\tau = m \lambda\), where \(\tau\) is the proper time and \(m\) satisfies
\begin{equation}
g_{\mu\nu}\dot{x}^\mu \dot{x}^\nu = -m^2 .
\end{equation}
For neutral particles, the values of \(m^2\) are \(1\), \(0\), and \(-1\) for timelike, null, and spacelike geodesics, respectively. The conjugate four-momenta of the particle are given by
\begin{equation}
    P_\mu = \frac{\partial{\mathcal{L}}}{\partial\dot{x}^\mu} = g_{\mu\nu}\dot{x}^\nu .
\end{equation} 
The Hamiltonian of the motion can be obtained using the Legendre transformation as follows,
\begin{equation}
 \mathcal{H} = P_\mu \dot{x}^\mu - \mathcal{L} = \frac{1}{2}g^{\mu\nu}P_\mu P_\nu .
\end{equation}
The action \( S \) for the particle is separable in the form
\begin{equation}
    S = \frac{1}{2}m^2\lambda - Et + L\varphi + S_r(r) + S_\theta(\theta) .
\end{equation}
Here, \( E \) and \( L \) are the conserved energy and angular momentum of the particle, respectively. In fact, these quantities are associated with the symmetry of the system, namely the time translation and the rotational symmetry (and hence associated with the Killing vectors \( t^a \) and \( \varphi^a \)). From Hamilton-Jacobi theory, we know that the conjugate momentum is related to the Hamilton-Jacobi action by \( P_\mu = \frac{\partial S}{\partial x^\mu} \). Therefore, we can write the conjugate momenta as
\begin{equation}\label{conjugatemom}
    \begin{split}
         P_t = -E, \qquad P_\varphi = L, \qquad P_r = \frac{\partial S}{\partial r}, \qquad P_\theta = \frac{\partial S}{\partial \theta}.
    \end{split}
\end{equation}
The Hamilton-Jacobi equation is given as follows,
\begin{equation}
\mathcal{H} = - \frac{\partial S}{\partial \lambda} .
\end{equation}
Substituting the value of the conjugate momenta and simplifying, we get
\begin{equation}
\begin{aligned}
\left(\frac{\partial S}{\partial \theta}\right)^2 + \left( aE\sin{\theta} - \frac{L}{\sin{\theta}} \right)^2 + m^2 a^2\cos^2{\theta} = &-\Delta\left(\frac{\partial S}{\partial r} \right) ^2 + \frac{[(a^2 + \ell^2 + r^2)E - aL]^2}{\Delta} \\ & - m^2(r^2 + \ell^2) .
\end{aligned}
\end{equation}
These equations are variable separable partial differential equations. Hence we can equate both the left-hand side and right-hand side to a constant, say \(\mathcal{K}\), such that
\begin{equation}
 \left(\frac{\partial S}{\partial \theta}\right)^2 = \mathcal{K} - \left( aE\sin{\theta} - \frac{L}{\sin{\theta}} \right)^2 - m^2 a^2\cos^2{\theta} ,
\end{equation}
\begin{equation}
 \Delta \left(\frac{\partial S}{\partial r} \right) ^2 = -\mathcal{K} + \frac{[(a^2 + \ell^2 + r^2)E - a L]^2}{\Delta} - m^2(r^2 + \ell^2) .
\end{equation}
Using \eqref{conjugatemom} for \( P_r \) and \( P_\theta \), we obtain
\begin{equation}
 \frac{d\theta}{d\lambda} = \dot{\theta} = \pm\frac{\sqrt{\Theta}}{\rho^2} ,
\end{equation}
\begin{equation}
 \frac{dr}{d\lambda} = \dot{r} = \pm\frac{\sqrt{\mathcal{R}}}{\rho^2} ,
\end{equation}
where,
\begin{equation}
 \Theta = \mathcal{K} - (L - aE)^2 - \cos^2{\theta}\left(\frac{L^2}{\sin^2{\theta}} + a^2(m^2 - E^2) \right) , 
\end{equation}
\begin{equation}
\mathcal{R}(r) = P(r)^2 - \Delta[(L - aE)^2 + m^2(r^2 + \ell^2) + \mathcal{Q}] ,
\end{equation}
\begin{equation}
 P(r) = E(r^2 + \ell^2 + a^2) - La .
\end{equation}
In the above equations, the constant \(\mathcal{Q}\) is the Carter constant, which is related to the separation constant \(\mathcal{K}\) as \(\mathcal{Q} = \mathcal{K} - (L - aE)^2\). Similarly, using \eqref{conjugatemom} for \( P_t \) and \( P_\varphi \), we obtain the Hamilton-Jacobi equations as follows
\begin{equation}
 -E = g_{tt}\dot{t} + g_{t\varphi}\dot{\varphi} ,
\end{equation}
\begin{equation}
 L = g_{\varphi\varphi}\dot{\varphi} + g_{t\varphi}\dot{t} .
\end{equation}
Simultaneously solving for \( \dot{t} \) and \( \dot{\varphi} \), we obtain
\begin{equation}
\dot{t} = \frac{a(L - aE\sin^2\theta)}{\rho^2} + \frac{r^2 + \ell^2 + a^2}{\rho^2\Delta} P(r) ,
\end{equation}
\begin{equation}
 \dot{\varphi} = \frac{(L - aE\sin^2\theta)}{\rho^2 \sin^2\theta}  + \frac{a}{\rho^2\Delta}P(r) .
\end{equation}

Having obtained the equations of motion for particles around the black hole, we now focus on particle collisions in the vicinity of the black hole horizon, specifically in the ergoregion. However, not every particle heading toward the black hole will enter the ergoregion. Depending on their angular momentum, some particles will get scattered, while others may move towards the event horizon. The latter are the particles that participate in collisions. We will consider test particles restricted to move on the equatorial plane, that is, \(\theta = \pi/2\). The equatorial plane is chosen because it simplifies our equations since Carter's constant \(\mathcal{Q}\) reduces to zero on the equatorial plane. Moreover, the ergoregion has maximum extension on the equatorial plane (see Fig. \ref{fig_ergo}). To find the range of angular momentum \(\ell\) for which a particle will reach the event horizon, we consider the radial motion of the particle, described by the equation
\begin{equation}
 \frac{1}{2}\left(\frac{\partial r}{\partial \lambda}\right)^2 + V_{\text{eff}} = 0 ,
\end{equation}
where \( V_{\text{eff}} \) is given by
\begin{equation}
 V_{\text{eff}} = \frac{[E(a^2 + r^2 + \ell^2) - aL]^2 - \Delta[m^2(r^2 + \ell^2) + (aE - L)^2]}{2(r^2 + \ell^2)^2} .
\end{equation}
The circular orbits are characterized by,
\begin{equation}
 V_{\text{eff}} = 0, \qquad \frac{dV_{\text{eff}}}{dr} = 0 .
\end{equation}
These two equations determine the maximum and minimum values of the angular momentum for a circular orbit. We have tabulated the values for both the extremal and non-extremal cases in Table \ref{tab2}. The range of angular momentum can be calculated for both photons and massive particles. Although photons have a broader range of angular momentum compared to massive particles, our study concentrates on the collision of massive particles, specifically focusing on cold dark matter particles \cite{Banados:2009pr}.

\begin{table}[t!]
\begin{tabular}{m{3em} m{3.5em} m{3.7em} m{3.5em} m{3.5em} || >{\centering} m{2em} m{3.7em} m{3.5em} m{3.5em}  }
 $l $  & $a_{\text{e}}$ &  $L_{\text{min}}$ & $L_{\text{max}}$ & $L_{\text{c}}$ & $a$ &  $L_{\text{min}}$ & $L_{\text{max}}$ & $L_{\text{c}}$\\
\hline
 $ 0 $ & 1  &  -4.82842  &  2  &  2  &  0.9 &  -4.75680  &  2.63245  &  3.19087   \\
 $ 0.1 $  & 1.00250  &  -4.83196  &  1.99998  &  1.99998  &  0.9 &  -4.75862  &  2.64031  &  3.20911   \\
 $ 0.2 $  & 1.01005  &  -4.84256  &  1.99969  &  1.99969  &  0.9 &  -4.76405  &  2.66334  &  3.26313   \\
 $ 0.3 $   & 1.02277  &  -4.86024  &  1.99832  &  1.99832  &  0.9 &  -4.77306  &  2.70004  &  3.35108   \\
 $ 0.4 $  & 1.04091  &  -4.88501  &  1.99422  &  1.99422  &  0.9 &  -4.78561  &  2.74842  &  3.47046   \\
 $ 0.5 $  & 1.06493  &  -4.91700  &  1.98382  &  1.98382  &  0.9 &  -4.80163  &  2.80635  &  3.61872   \\
 $ 0.60057 $   & 1.09596  &  -4.95681  &  1.95755  &  1.95755  &  0.9 &  -4.82115  &  2.87223  &  3.79462   \\
 $ 0.70711 $  & 1.13975  &  -5.00927  &  1.79779  &  1.79779  &  0.9 &  -4.84545  &  2.94850  &  4.00818   \\
\hline
\end{tabular}
\caption{Maximum \(L_{\text{max}}\) and minimum \(L_{\text{min}}\) angular momentum of massive particles approaching an extremal black hole (left panel) and a non-extremal black hole (right panel).} \label{tab2}
\end{table}

\begin{figure}[t!]
    \centering
    \includegraphics[width=\textwidth]{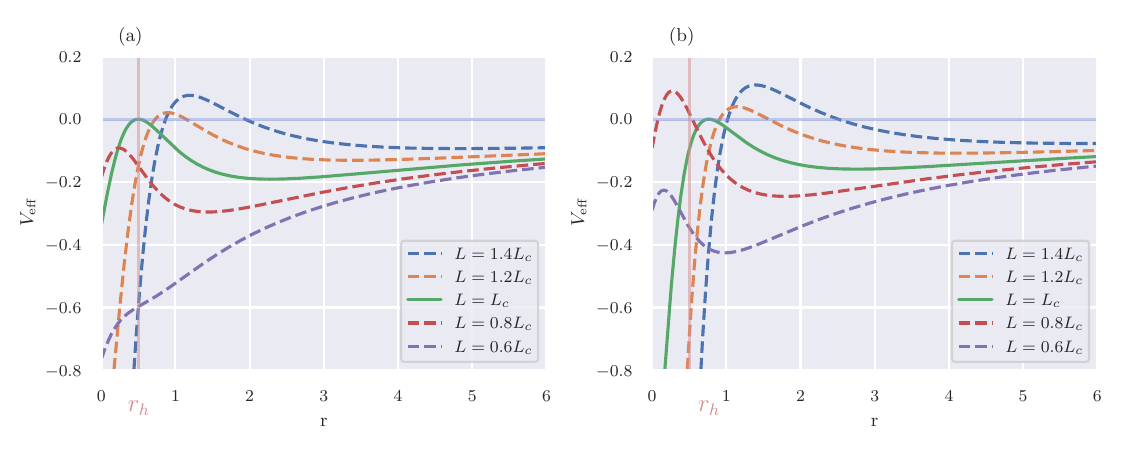}
    \caption{(a) Effective potential for the ultraextremal black hole ($a_{\text{e}}=1.13975$, $\ell  _{\text{e}}=0.70711$, $r_{\text{h}}=0.5$ and $\ell = 1.79778$). (b) Effective potential for an extremal black hole with a null throat ($a_{\text{e}} = 1.09596$, $\ell  _{\text{e}}= 0.60057$, $r_{\text{h}} = 0.76393$ and $L_{\text{c}} = 1.95755$).}
   \label{fig_veff_ultra}
\end{figure}

\begin{figure}[t!]
    \centering
    \includegraphics[width=\textwidth]{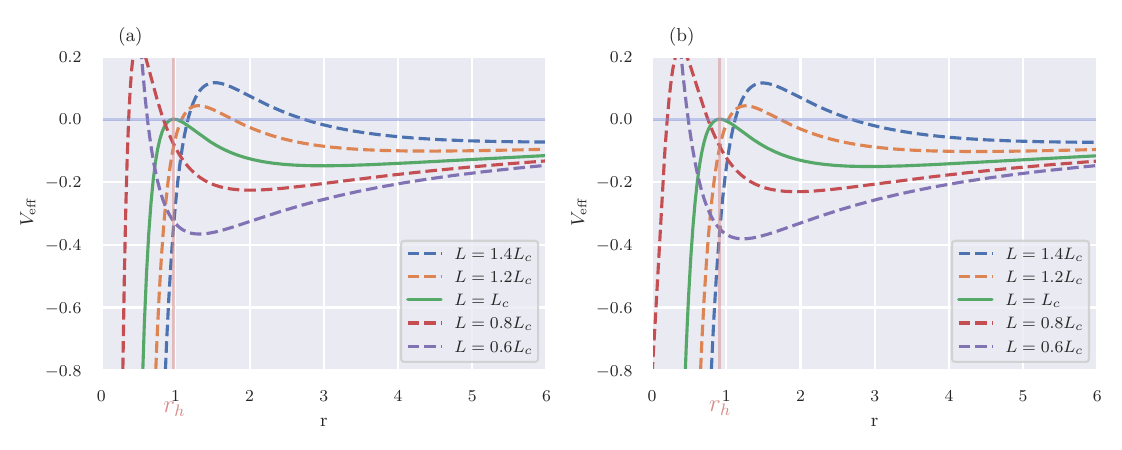}
    \caption{Effective potential for the extremal black holes. (a) $a_{\text{e}} = 1.01005$, $\ell = 0.2$, $r_{\text{h}} = 0.97958$ and $L_{\text{c}} = 1.99968$.  (b) $a_{\text{e}} = 1.04091$, $\ell = 0.4$, $r_{\text{h}} = 0.91231$ and $L_{\text{c}} = 1.99421$.}
   \label{fig_veff_next}
\end{figure}

Given that the geodesics are time-like for massive particles, \(\dot{t} \geq 0\), we have
\begin{equation}
    \frac{a(L - aE\sin^2\theta)}{\rho^2} + \frac{r^2 + \ell^2 + a^2}{\rho^2\Delta} P(r) \geq 0 .
\end{equation}
Taking the limiting case, setting \(\theta = \pi/2\) and re-arranging the equation, we get
\begin{equation}
    L = \frac{\left(a^2\Delta - (r^2 + \ell^2 + a^2)^2\right)}{a(\Delta - (r^2 + \ell^2 + a^2))} E .
\end{equation}
In the limit \(r \rightarrow r_{\text{h}}\), the equation reduces to
\begin{equation}\label{lceqn}
L = \frac{r_{\text{h}}^2 + \ell^2 + a^2}{a} E = \frac{E}{\Omega_{\text{h}}}  \equiv L_{\text{c}}  
\end{equation}
where $\Omega_{\text{h}}$ is the angular velocity of the black hole on the horizon,
\begin{equation}
    \Omega_{\text{h}} = \frac{a}{r_{\text{h}}^2 + a^2 + \ell^2} ,
\end{equation}
and $L_{\text{c}}$ is the critical angular momentum of the particle. Particles that approach the black hole with angular momentum exceeding the critical value will be scattered away before it reaches the horizon. The particles with angular momentum within the range \((L_{\text{min}} , L_{\text{max}})\) will be absorbed by the black hole. These behaviors of the particle motion can be easily understood by analyzing the effective potential (Fig. \ref{fig_veff_ultra} and Fig. \ref{fig_veff_next}). For all $L < L_{\text{c}}$, the effective potential is always negative, resulting in bounded motion. However, for angular momentum greater than the critical value, the particle encounters an effective potential barrier as it approaches the black hole, leading to unbounded motion. This behavior remains consistent across all extremal black hole cases. In some instances, potential bumps exist for $L < L_{\text{c}}$ inside the horizon, but these do not influence the particle motion outside the horizon. 

\section{Particle Collisions near the Black Hole Horizon}\label{sec3}

In this section, we consider the collision of particles near the black hole horizon. We assume that the particles are initially at rest at infinity, so that the collisional energy is solely due to gravitational acceleration. The particles are assumed to have different angular momenta, where the range is specified by \(L_{\text{min}}\) and \(L_{\text{max}}\) obtained in the previous section. We consider the particle collision near the horizon at some radius \(r\) and then take the limiting case \( r \to r_{\text{h}} \). For simplicity, the particle collision is considered on the equatorial plane \(\theta = \pi/2\) (hence \(\dot{\theta} = 0\)). 

The expression for calculating the center-of-mass energy of a collision in curved spacetime can be obtained as follows. When two particles collide, they occupy the same spacetime-point. An observer in the center-of-mass frame is defined as having a four-velocity parallel to the sum of the four-momenta \( p_1^\mu \) and \( p_2^\mu \) of the particles. The energy measured by this observer is called the center-of-mass energy, which is given by \cite{Harada:2014vka},
\begin{equation}
    E_{\text{CM}}^2 = -g_{\mu\nu}(p_1^\mu + p_2^\mu)(p_1^\nu + p_2^\nu) . 
\end{equation}
Substituting \( p^\mu = m \dot{x}^\mu \), the above equation can be rewritten for two identical particles of mass \( m \) as follows,
\begin{equation}
    E_{\text{CM}}^2 = -m^2 g_{\mu\nu}(\dot{x}_1^\mu + \dot{x}_2^\mu)(\dot{x}_1^\nu + \dot{x}_2^\nu) .
\end{equation}
Using the normalization condition of the four-velocity \( u^{\mu} \), \( u^{\mu}u_{\mu} = -1 \), we obtain \cite{Banados:2009pr},
\begin{equation} \label{comenergy}
\frac{E_{\text{CM}}^2}{2m^2} = \left(1 - g_{\mu\nu}\dot{x}_1^{\mu}\dot{x}_2^{\nu}\right).
\end{equation}
From the equations of motion derived in the previous section, the four-velocity of the particles can be written as,
\begin{equation} 
\dot{x}_i^{\mu} = \left(\frac{a(L_i - aE_i)}{r^2 + \ell^2} + \frac{r^2 + \ell^2 + a^2}{(r^2 + \ell^2)\Delta}P(r), \frac{\sqrt{\mathcal{R}}}{r^2 + \ell^2}, 0, \frac{(L_i - aE_i)}{r^2 + \ell^2} + \frac{aP(r)}{(r^2 + \ell^2)\Delta} \right).
\end{equation}
Using this and substituting \(E_i = 1\) in \eqref{comenergy}, we get,
\begin{equation}\label{ecmeqn}
\begin{aligned}
\Xi &\equiv \frac{E_{\text{CM}}^2}{2m^2} \\ &= \frac{\left(a^2-a L_1+\ell^2+r^2\right) \left(a^2-a L_2+\ell^2+r^2\right)+\Delta  \left(\ell^2+r^2-(a-L_1) (a-L_2)\right) - X_1 X_2 }{\Delta  \left(\ell^2+r^2\right)}
\end{aligned} 
\end{equation}
where,
\begin{equation} 
X_i = \sqrt{\left(a^2 - aL_i + \ell^2 + r^2\right)^2 - \Delta (m^2 (\ell^2 + r^2) + (a - L_i)^2)}.
\end{equation}
The above equation remains unchanged if \(L_1\) and \(L_2\) are interchanged. This implies that the center-of-mass energy \(E_{\text{CM}}\) is invariant under the exchange of particles (since we consider identical particles). We also note that the value of \(E_{\text{CM}}\) deviates from the Kerr black hole case as it depends on the LQG parameter \(\ell\), in addition to the black hole spin. This implies that the particle acceleration in the LQG black hole background is influenced by \(\ell\). We examine the characteristics of the center-of-mass energy as the radial coordinate \( r \) nears the event horizon \( r_\text{h} \) of the black hole. By taking the limit as \( r \to r_\text{h} \) in the extremal scenario (applying L'Hôpital's rule twice), we find

\begin{equation}\label{ecmlimit}
\begin{aligned}
\frac{E_{\text{CM}}^2}{2m^2} &= 9.0151 - A(L_1, L_2)  - \frac{0.3043 \cdot B(L_1) \cdot B(L_2)}{C(L_1) \cdot C(L_2)}  + \frac{C(L_2) \cdot D(L_1)}{C(L_1) \cdot F(L_1)} + \frac{C(L_1) \cdot D(L_2)}{C(L_2) \cdot F(L_2)},
\end{aligned}
\end{equation}
where
\[
\begin{aligned}
A(L_1, L_2) &= 1.2672 \, L_1 + L_1 \left(1.0490 - 1.0077 \, L_2\right) - 0.2182 \, L_2, \\
B(L_z) &= 7.5751 - 3.7985 \, L_z, \\
C(L_z) &= \sqrt{4.3090 + L_z \left(-4.3215 + 1.0835 \, L_z\right)}, \\
D(L_z) &= -5.8901 + L_z \left(6.7749 + L_z \left(-0.3475 + \left(-1.7915 + 0.5039 \, L_z\right) L_z\right)\right), \\
F(L_z) &= 3.9769 + L_z \left(-3.9884 + \, L_z\right).
\end{aligned}
\]
Here, we have taken \( \ell = 0.4 \) and \( M = 1 \), along with the corresponding values for the angular momentum and horizon radius of the extremal black hole. Expression \eqref{ecmlimit} diverges when one of the particles approaches the black hole with critical angular momentum \( L_{\text{c}} \), as \( C(L_z) \) becomes zero in such cases. An analytical expression for the limiting values of the center-of-mass energy can be derived for any other value of \( \ell \). However, it should be noted that this divergence occurs only for extremal black holes.

\begin{figure}[t!]
    \centering
    \includegraphics[width=\textwidth]{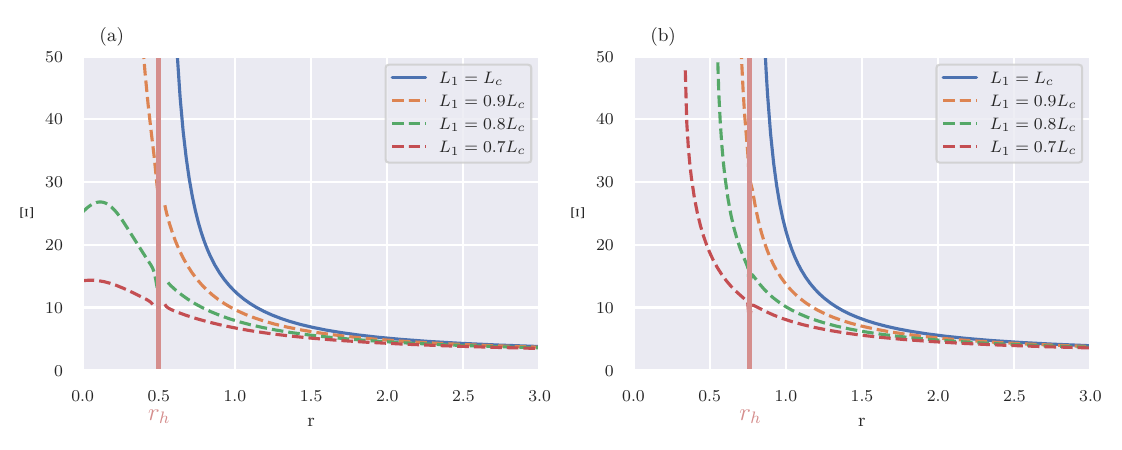}
    \caption{(a) Center-of-mass energy for the ultraextremal black hole ($a_{\text{e}}= 1.13975$,  $\ell_{\text{e}}=0.70711$, $r_\text{he}=0.5$, $L_\text{c}=1.79778$ and $L_2=L_\text{min}=-5.00926$). (b) Center-of-mass energy for an extremal black hole with a null throat ($a_{\text{e}}=1.09596$, $\ell =0.60057$, $r_\text{he}=0.76393$, $L_\text{c}=1.95755$ and $L_2=L_\text{min}=-4.95681$).}
    \label{com_fig_ultra}
\end{figure}

\begin{figure}[t!]
    \centering
    \includegraphics[width=\textwidth]{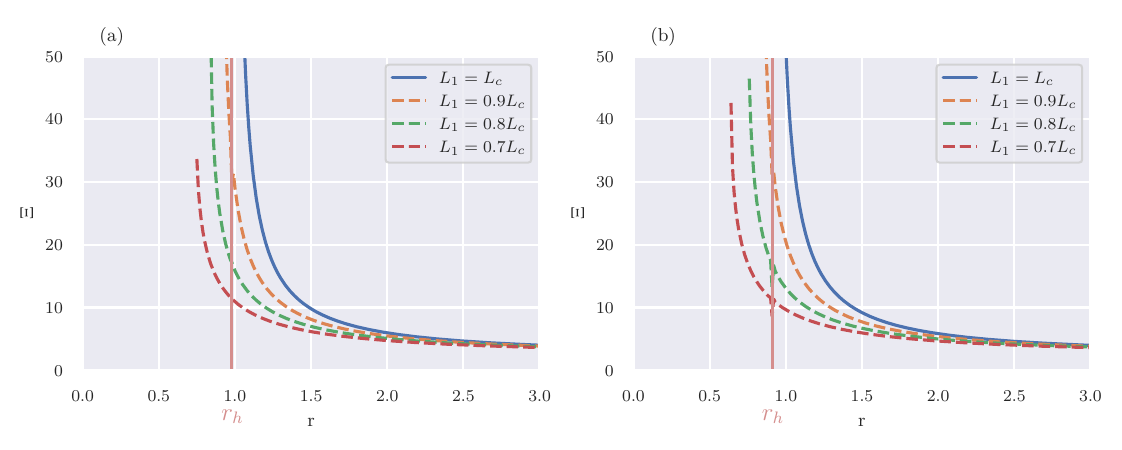}
    \caption{Center-of-mass energy for the extremal black hole. (a) $\ell=0.2$, $a_{\text{e}}=1.01005 $, $r_\text{he}=0.97958$, $L_\text{c}=1.99968$ and $L_2=L_\text{min}=-4.84256$. (b) $\ell=0.4$, $a_{\text{e}}=1.04091 $, $r_\text{he}=0.91231$, $L_\text{c}=1.99421$ and $L_2=L_\text{min}=-4.88501$.}
    \label{com_fig_ext}
\end{figure}

\begin{figure}[t!]
    \centering
    \includegraphics[width=\textwidth]{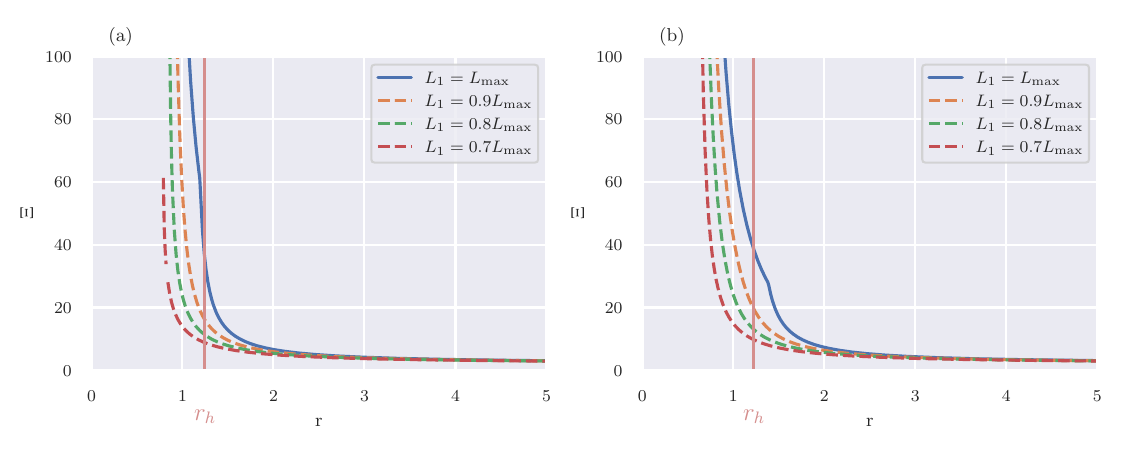}
    \caption{Center-of-mass energy for the non-extremal black hole. (a) $a=1$, $\ell=0.2$, $r_\text{h}=1.12432$, $L_\text{max}=2.20028$ and $L_2=L_\text{min}=-4.83548$. (b) $a=1$, $\ell=0.4$, $r_\text{h}= 1.22264$, $L_\text{max}=2.40120$ and $L_2=L_\text{min}=-4.85651$. }
    \label{com_fig_nonext}
\end{figure}

In Fig. \ref{com_fig_ultra}, \ref{com_fig_ext} and \ref{com_fig_nonext}, we study the behavior of \(E_{\text{CM}}\) as a function of the radial coordinate \(r\) in the vicinity of the event horizon \(r_{\text{h}}\), for extremal and non-extremal cases. The angular momenta of the particles are taken within the allowed range \(L_{\text{min}}\) and \(L_{\text{max}}\) so that they always approach the event horizon. Only particles possessing critical angular momentum can reach the horizon. However, it is important to note that the critical angular momentum, determined by Eq.~\eqref{lceqn}, falls outside the range \((L_{\text{min}}, L_{\text{max}})\) for non-extremal black holes, but it is within this range for extremal black holes. Without loss of generality, we set $m=1$ in our analysis.

For extremal black holes, as shown in Fig.~\ref{com_fig_ultra} and \ref{com_fig_ext}, it is clear that the center-of-mass energy \(E_{\text{CM}}\) remains finite when the angular momenta of the incoming particles are below the critical value, and it diverges when one of the particles has critical angular momentum. Although this result is similar to that of the Kerr black hole, the extremal spin in the presence of the LQG parameter can exceed the Kerr case, enhancing the potential to probe Planck-scale physics through particle acceleration. However, it is important to note that the extremal case is an idealization, as it takes an infinite amount of time for a particle to reach the horizon. Therefore, we expect that near-extremal cases can act as realistic particle accelerators, even though the time required is very long.

We investigate the behavior of \(E_{\text{CM}}\) for non-extremal black holes in Fig.~\ref{com_fig_nonext}, where it attains finite and sufficiently high values at the horizon in all cases. This behavior is explored through a numerical approach, where we calculate the center-of-mass energy for a near-extremal black hole with angular momentum \(a = a_\text{e} - \epsilon\), where \(\epsilon\) is a small parameter characterizing a slight deviation from the extremal case \(a = a_\text{e}\). In our numerical computations, we use \(L_1 = L_\text{min}\) and \(L_2 = L_\text{max}\) in the expression for the center-of-mass energy \eqref{ecmeqn} and take the limit \(r \to r_\text{h}\), where \(L_\text{min}\), \(L_\text{max}\), and \(r_\text{h}\) are numerically determined for \(a = a_\text{e} - \epsilon\). The results, presented in Table~\ref{tab:ecm_tune}, consider smaller values of \(\epsilon\) and various \(\ell\) values. The data in Table~\ref{tab:ecm_tune} clearly demonstrate that the center-of-mass energy \(E_\text{CM}\) increases gradually as the black hole approaches the maximally spinning limit (\(\epsilon \to 0\)). For a fixed value of \(\epsilon\), \(E_\text{CM}\) rises with increasing values of the LQG parameter \(\ell\). This behavior contrasts with the Kerr-Newman case \cite{Wei:2010vca}, where \(E_\text{CM}\) decreases as the charge increases, but aligns with the Kerr-Sen black hole scenario \cite{Wei:2010gq}, in which \(E_\text{CM}\) increases with increasing charge. Additionally, when \(\ell = 0\), the spacetime corresponds to a Kerr black hole, and the results in Table~\ref{tab:ecm_tune} are consistent with those reported in Ref.~\cite{Jacobson:2009zg}.

\begin{table}[t!]
\begin{tabular}{m{3.5em} m{5.5em} m{5.5em} m{5.5em} m{5.55em} m{5.5em} m{5.5em} m{5.5em} m{5.5em}}
 $\ell$ & $\epsilon = 0.1$ & $\epsilon = 0.05$ & $\epsilon = 0.01$ & $\epsilon = 0.001$ & $\epsilon = 0.0001$ \\
\hline
 $ 0 $    & 6.90064 & 8.24403 & 12.5353 & 22.6342 & 40.4856 \\
 $ 0.1 $  & 6.90387 & 8.24916 & 12.5464 & 22.6584 & 40.5315 \\
 $ 0.2 $  & 6.91397 & 8.26531 & 12.5816 & 22.7356 & 40.6787 \\
 $ 0.3 $  & 6.93246 & 8.29503 & 12.6475 & 22.8826 & 40.9597 \\
 $ 0.4 $  & 6.96237 & 8.34370 & 12.7590 & 23.1386 & 41.4553 \\
 $ 0.5 $  & 7.00982 & 8.42251 & 12.9495 & 23.5977 & 42.3625 \\
 $ 0.6 $  & 7.08872 & 8.55867 & 13.3125 & 24.5610 & 44.3473 \\
 $ 0.7 $  & 7.24973 & 8.86388 & 14.3815 & 28.7519 & 55.5305 \\
\hline
\end{tabular}
\caption{ The center-of-mass energy per unit rest mass $ E_{\text{CM}} /m$ with spin $ a = a_{\text{e}} - \epsilon $, and, $ L_1 = L_{\text{max}}, L_2 = L_{\text{min}} $.}
\label{tab:ecm_tune}
\end{table}

Lastly, we briefly assess the tolerance level of the critical angular momentum $ L_{\text{c}}$ in the BSW mechanism. For simplicity, we assume \( L_1 = L_\text{c} - \delta L \) and \( L_2 = 0 \), which simplifies Eq. \eqref{ecmlimit} to
\[
\frac{ E^2_{\mathrm{CM}}}{2m ^2} \approx 9.60695\times 10^{17} \delta L+\frac{3.82061\times 10^{18}}{\delta L}+1.68323\times 10^{17}.
\]
Assuming the rest mass of the colliding particle is \( 1 \, \text{GeV} \), comparable to a neutron, and setting \( E_{\mathrm{CM}} = 10^{20} \, \text{eV} \), we find
\[
\delta L \approx 3.82061 \times 10^{-22}
\]
which represents the tolerance for the critical angular momentum \( L_{\text{c}} \).

\section{Discussions}\label{sec4}

In this article, we have shown that a spinning LQG black hole can act as a particle accelerator. Such a black hole solution is particularly interesting because quantum effects resolve the classical singularity problem, and the limits on the spin of extremal black holes are more relaxed compared to the Kerr black hole. The deviation in spacetime properties compared to the Kerr black hole, including the multi-horizon structure, is induced by the LQG parameter $\ell$. Although black hole solutions can exist for all values of $\ell$, extremal solutions are guaranteed only within a specific parameter range $\ell \leq \ell_\text{e}$. We have studied the horizon structure and ergoregion in the domain $0 \leq \ell \leq \ell_\text{e}$ for both extremal and non-extremal cases, and observed that they are significantly different from the Kerr spacetime. For a fixed value of $\ell$, the extremal solution is characterized by the corresponding extremal spin $a_\text{e}$, which includes an ultraextremal black hole, defined by $(a_{\text{e}} = 1.13975, \ell  _{\text{e}}= 0.70711)$, and an extremal black hole with a null throat, defined by $(a_{\text{e}} = 1.09596, \ell = 0.60057)$.

Utilizing the BSW mechanism, we have explored the center-of-mass energy characteristics for two particles colliding in the equatorial plane of a spinning LQG black hole. Astrophysical black holes may be surrounded by relic cold dark matter density spikes, and it is widely recognized that cold dark matter does not interact electromagnetically with other forms of matter. Therefore, our study focuses on the collision of two uncharged, massive particles. We analyze particle motion by solving the geodesic equations, which allow us to determine the range of angular momentum enabling a particle to reach the black hole. This range corresponds to the effective potential, which either permits or restricts the particle's approach. For particles within the allowed range, we examine collisions near the black hole. Our findings indicate that when one of the particles possesses critical angular momentum, the center-of-mass energy becomes exceedingly high in the spacetime of an extremal black hole. Conversely, for non-extremal black holes, the center-of-mass energy remains finite, even when the incoming particle has critical angular momentum. The behavior of BSW mechanism is influenced by the LQG parameter $\ell$, which also governs the horizon structure.

The observation that center-of-mass energy diverges during particle collision is particularly intriguing from an astrophysical perspective. Supermassive black holes, such as those located at the center of the giant elliptical galaxy M87$^\star$ and at the center of our galaxy, Sgr A$^\star$, might accelerate particles through this mechanism, potentially linking it to the observed ultra-high-energy cosmic rays, which can reach up to $10^{20}$ eV. These collisions could also produce exotic massive particles, offering an avenue for further investigation in a realistic setting as an extension of our study. Consequently, we believe that our research could be connected to observational phenomena aimed at probing Planck-scale physics. A potential direction for future work involves calculating the efficiency of energy extraction in the collisional Penrose process for spinning LQG black holes. From table \ref{tab:ecm_tune}, we observe an enhancement in center-of-mass energy, which could result in increased efficiency compared to the Kerr black hole.

\section{Acknowledgment}
ANK's research was supported by the Croatian Science Foundation Project No. IP-2020-02-9614 \textit{Search for Quantum spacetime in Black Hole QNM spectrum and Gamma Ray Bursts}.

\bibliography{BibTex.bib}
\end{document}